\documentclass[aps,prl,twocolumn,showpacs]{revtex4}

\usepackage{epsfig}

\usepackage{dcolumn}

\usepackage{bm}

\newcommand{\bk}{{\bf k}}

\newcommand{\pd}{$P^{ave}_d$ }

\newcommand{\npi}{$n({\bf k}=(\pi,0))$ }

\begin{document}








\title{Enhancement of Pairing Correlation by $t'$ in the Two-Dimensional Extended $t-J$ Model}










\author{C.~T. Shih$^{1}$, T.~K. Lee$^{2}$, R. Eder$^{3}$, C.-Y. Mou$^{4}$ and Y.~C. Chen$^{1}$}

\address{$^1$Department of Physics, Tunghai University, Taichung,
Taiwan\\
$^2$Institute of Physics, Academia Sinica, Nankang,
Taiwan\\
$^3$Institut f\"ur Festk\"orperphysik, Forschungszentrum
Karlsruhe, Germany\\
$^4$Department of Physics, National Tsing Hua University,
Hsinchu, Taiwan}

\date{\today}

\begin{abstract}
We investigate the effects of the next-nearest-neighbor ($t'$) and
the third-nearest-neighbor ($t''$) hopping terms on
superconductivity correlation in the 2D hole-doped extended
$t-J$ model based on the variational Monte-Carlo, mean-field
calculation, and exact diagonalization method. Despite
of the diversity of the methods employed, the results all point to
a consistent conclusion: While the $d-$wave SC correlation is
slightly suppressed by $t'$ and $t''$ in underdoped regions, it is
greatly enhanced in the optimal and overdoped regions. The optimal
$T_c$ is a result upon balance of these two opposite trends.
\end{abstract}

\maketitle

Right after the discovery of high temperature superconductors, the
two-dimensional (2D) $t-J$ model has been proposed to provide the
mechanism of superconductivity (SC)\cite{anderson87}. This idea quickly gained
momentum when variational calculations showed that the doping
dependence of pairing correlation\cite{gros88,zhang88} and the
phase diagram of antiferromagnetic (AFM) phase and  SC
\cite{lee88} seem to agree with experimental results fairly well.
However, recently calculations\cite{shih98} beyond the variational
method have challenged the notion that pure 2D $t-J$ model without
including other interactions is enough to explain the high values
of $T_c$. Although this issue is yet to be
settled\cite{sorella02,lee02,sorella02b}, there are
results by band-structure calculations\cite{raimondi96,pavarini01}
and experimental analysis\cite{fujimori} that hopping beyond
nearest neighbors is essential to raise  $T_c$. In fact they found
highest $T_{c,max}$ for different monolayer cuprates strongly
correlates with $t'/t$, where $t'$ is the second nearest-neighbor
hopping amplitude. This contradicts with previous
results\cite{white99,martins01} of exact calculations
 that  for the hole doped systems, introducing $t'$
into the t-J model will actually reduce pairing. Although the last
results are studied for systems doped only with a few  number of
holes, all these conflicting results raised a very serious
challenge to the $t-J$ type models.

There are many experimental and theoretical results to support the
presence of $t'$ and possibly also $t''$, the third nearest
neighbor hopping, in cuprates. The topology of large Fermi surface
seen by ARPES\cite{damascelli03,ino02} and
the change of sign of Hall coefficient as a function of
doping\cite{clayhold89}
 can best be understood in the
presence of $t'$ and $t''$. The single hole dispersion observed by
ARPES and the difference between hole- and electron-doped
systems\cite{lee03} also support the presence of these terms.

In this letter, we will use variational approach supplemented by
slave-boson mean-field (MF) calculations and exact diagonalization (ED)
 method to show that the
presence of $t'$ is indeed important for enhancing pairing beyond the
underdoped regime. The largest values of pairing correlation
obtained are proportional to $t'/t$ up to $t'/t=-0.3\sim -0.4$. In
addition we will show that the decrease of pairing correlation  at
very large hole density is related to the change of Fermi surface
topology. The conflicting results between theories and experiments
discussed above are naturally resolved within the extended $t-J$
model.

The Hamiltonian of the extended $t-J$ model is
\begin{eqnarray}
&H&=H_t+H_J=\\
&-&\sum_{ij}t_{ij}(\tilde{c}^\dagger_{i,\sigma}\tilde{c}_{j,\sigma}
+ H.C.)+J\sum_{<i,j>}({\bf{S_i}\cdot
S_j}+\frac{1}{4}n_in_j)\nonumber \label{e:tjm}
\end{eqnarray}
where $t_{ij}=t$, $t'$, and $t''$ for sites $i$ and $j$ are
nearest, next nearest, and the third nearest neighbors,
respectively, and $t_{ij}=0$ for longer distance.
$\tilde{c}_{i,\sigma}=(1-n_{i,-\sigma})c_{i,\sigma}$, satisfies
the no-double-occupancy constraint. In our notation for hole doped
materials, $t'/t$ is negative while $t''=-t'/2$ most of the time.

The trial wave function (TWF) used in this study is the
$d_{x^2-y^2}$ resonating-valence-bond wave function
\begin{eqnarray}
\mid \Psi\rangle =
P_G\prod_k(u_k+v_kc^\dagger_{k,\uparrow}c^\dagger_{-k,\downarrow})\mid
0\rangle
\end{eqnarray}
with
$u_k/v_k=\Delta_k/(\epsilon_k+\sqrt{\epsilon_k^2+\Delta_k^2})$,
$\epsilon_k=-2t(cosk_x+cosk_y)-4t'_vcosk_xcosk_y-2t''_v(cos2k_x+cos2k_y)-\mu$,
and $\Delta_k=2\Delta(cosk_x-cosk_y)$. Here the projection
operator $P_G$ enforces the constraint of one electron per site.
In addition to $\Delta$ and $\mu$, we have
included two important variational parameters $t'_v$ and $t''_v$ which
determine
the Fermi surface topology.

The d-wave pair-pair correlation $P_d({\bf R})$ is defined as
$\frac{1}{N_s}\langle\sum_i\Delta^\dagger_{\bf R_i}\Delta_{\bf
R_i+R}\rangle$, where $\Delta_{\bf R_i}=c_{{\bf
R_i}\uparrow}(c_{{\bf R_i+\hat{x}}\downarrow}+c_{{\bf
R_i-\hat{x}}\downarrow}-c_{{\bf R_i+\hat{y}}\downarrow}-c_{{\bf
R_i-\hat{y}}\downarrow})$. The long range part of $P_d({\bf R})$
is a flat plateau for nonzero $\Delta$, and we define $P^{ave}_d$
as the averaged value of the $\mid{\bf R}\mid > 2$ part of
$P_d({\bf R})$ to estimate the strength of SC of the system.

\begin{figure}[here]
\rotatebox{-90}{\includegraphics[width=2.4in]{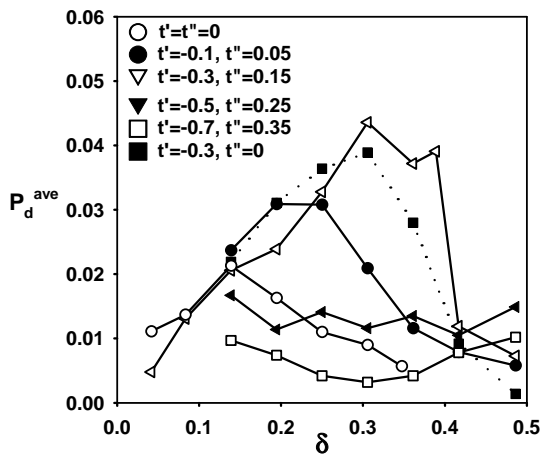}}
\caption{$P^{ave}_d$ for $J/t=0.3$ with several $t'$ and $t''$ for
a hole-doped $12\times 12$ lattice.} \label{f:phase_144v}
\end{figure}

Fig.\ref{f:phase_144v} shows the variational Monte Carlo (VMC)
results of $P^{ave}_d$ for several $t'$ and $t''$ for different
hole densities $\delta$ with $J/t=0.3$ in a $12\times 12$ lattice.
$P^{ave}_d$ for the $t-J$ model (open circles) has the
``dome-like'' shape which is similar to the experimental results
of $T_{c}$ versus doping. However, this could be an artifact of
the variational study which, we believe, overestimates the order
parameters and will be largely suppressed when we go beyond
variational calculation as shown in our previous
study\cite{shih98}. The most suprising result is that when $t'$ is
included, $P^{ave}_d$ changes dramatically. For the overdoped
regime, $P^{ave}_d$ is greatly enhanced by almost one order of
magnitude for $t'/t=-0.3$ and $t''/t=0.15$. The SC region extends
to $\delta\sim 0.4$ and the peak of the superconducting dome is at
$\delta\sim 0.3$ and the magnitude of the maximal \pd is about 2.5
times larger than for $t'=t''=0$. In the underdoped region, \pd is
almost unchanged or very slightly suppressed. Another thing to
note is that beyond the value of $t'/t=-0.3\sim-0.4$ $P^{ave}_d$
is no longer enhanced.

\begin{figure}[here]
\rotatebox{-90}{\includegraphics[width=2.3in]{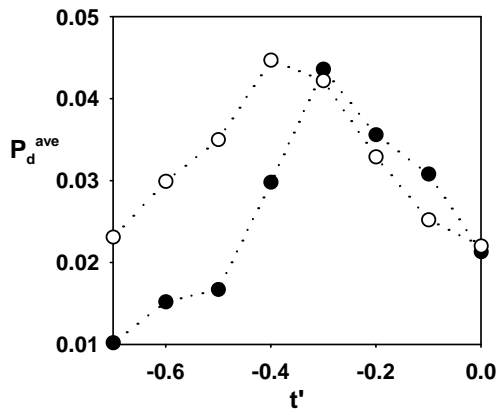}}
\caption{Maximal \pd for different $t'$ with $t''=-t'/2$ for
$8\times 8$ (open circle) and $12\times 12$ (full circle)
lattices.} \label{f:144t2}
\end{figure}

Fig.\ref{f:144t2} plots the maximal possible value of \pd for all doping density
as a function of $t'$. The
maximal \pd is proportional to $t'$ in the range $0\ge t'\ge
-0.3\sim-0.4$. Beyond these values pairing is no longer enhanced.
Coincidentally these values are about the same  value of $t'/t$
for mercury cuprates as estimated by Pavarini {\em et al.}\cite{pavarini01}
but much larger than what was reported in Ref.\cite{raimondi96}.
Among all the cuprate series, mecury
cuprate maintains the record of having highest $T_c$ for almost a decade.

The above result resolves the discrepancy between previous
denisty-matrix-renormalization-group (DMRG)
studies \cite{white99,martins01} and the band structure
calculation \cite{pavarini01}. As DMRG studies were concerned with underdoped
region
while the highest  $T_{c,max}$ examined by the band structure calculation
 certainly depends on
optimal and overdoped regions.

\begin{figure}[here]
\rotatebox{-90}{\includegraphics[width=2.7in]{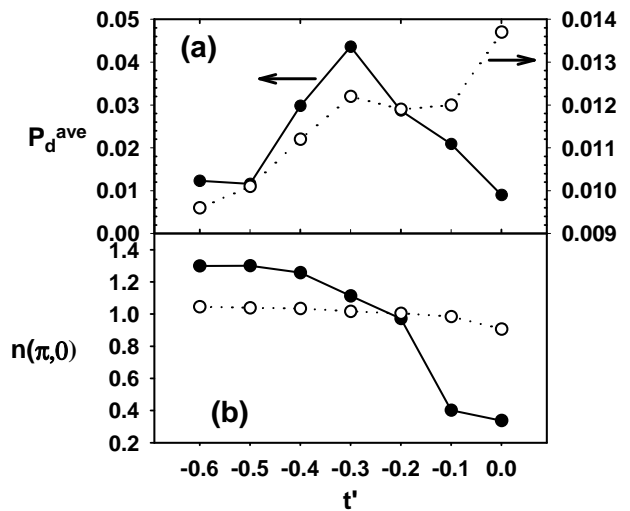}}
\caption{(a)\pd and (b) \npi vs. $t'$ for $\delta=0.31$ (full
circles) and $0.083$ (open circles), $J/t=0.3$, $t''=-t'/2$.}
\label{f:nk_pd}
\end{figure}

The different effects of $t'$ and $t''$ on \pd between
overdoped and underdoped regions are related to the
 shape of Fermi surface.
 Fig.\ref{f:nk_pd} shows the relations of \pd and
\npi versus $t'$ for $\delta=0.31$ and $0.083$. For the overdoped
 case ($\delta=0.31$), as $-t'$ increases up to $0.3$,  \npi increases
from less than $0.4$ to larger than $1$ and \pd also increases sharply from
less than $0.01$ to larger then $0.04$.
 Since $d-$wave SC gap
is largest at ${\bf k}=(\pi,0)$,
 occupation of this ${\bf k}$ states by electrons
 enhances \pd.
 On the other hand for the
underdoped case, \npi is almost unchanged because the
occupation \npi is already quite  large ($>0.9$) for $t'=t''=0$
and the effect of Fermi surface becomes unimportant.
The slight suppression of \pd
may be due to the destructive interference mechanism of the
pair-hopping as suggested by Martins {\em et al.}\cite{martins01}.

The decrease of \pd for $-t'\ge 0.4$ in the overdoped regime such
as $\delta=0.31$
 is also likely the consequence of the change of  the
Fermi surface. \npi is almost  saturated at $-t'=0.4$ and remains
unchanged for larger $-t'$. It is not difficult to recognize that as $-t'$
becomes much larger than $t$, electrons will occupy in separate
regions around $\bk=(\pm\pi,0)$ and $\bk=(0,\pm\pi)$. Hence the
Fermi surface becomes disjoint pieces. Although at $-t'/t=0.4$,
the Fermi surface is still connected but this tendency is already
observed. The density of states begins to decrease and this is
probably the reason for the suppression of pairing beyond
$-t'/t'\ge 0.4$.

\begin{figure}[here]
\rotatebox{-90}{\includegraphics[width=2.3in]{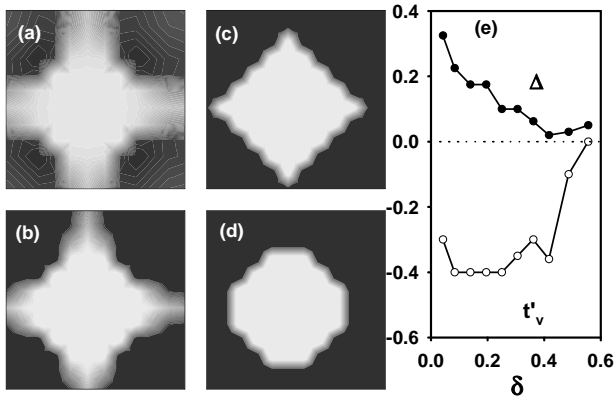}}
\caption{Fermi surface of $\delta=$ (a)0.19, (b)0.31 (c)0.42 (d)
0.49 for $12\times 12$ lattice. (e)optimal parameters $t'_v$
(squares) and $\Delta$ (circles).} \label{f:para144}
\end{figure}

Fig.\ref{f:para144} shows the Fermi surface and optimal parameters
$\Delta$ and $t'_v$ as a function of doping density for a
$12\times 12$ lattice with $J/t=0.3$, $t'=-0.3$ and $t''=0$. In
(a-d) the white region denotes $n({\bf k})\geq1.2$ and dark region
for  $n({\bf k})\leq0.5$. The density with maximal \pd is near
$\delta_{opt}=0.31$ as shown by the solid square in
Fig.\ref{f:phase_144v}. The shapes of the Fermi surfaces are very
different for $\delta>\delta_{opt}$ and $\delta<\delta_{opt}$
cases. Fig.\ref{f:para144}(e) shows that in the  region
$0.2<\delta<0.3$ although $\Delta$ becomes  smaller, $-t'_v$ is
still quite large and pairing is further enhanced. Doping beyond
$\delta>0.3$, $-t'_v$ begins to decrease quickly. This gives very
low electron occupation at  $(\pi,0)$ as shown in
Fig.\ref{f:para144}(c) and (d), then the pairing is reduced. This
result shows that the
enhancement of pairing by including $t'$ is not due to larger
$\Delta$ but from the deformation of the Fermi surface instead.

Although we have emphasized the
particular correlation between d-wave SC gap and electron
occupation at ${\bf k}=(\pi,0)$ as the reason for enhancement of
pairing, another familiar effect may also have played a role.
It is well known that
$t'$ will shift the van Hove singularity\cite{newns} in density of states,
but it is always around  ${\bf k}=(\pi,0)$. Hence it could be that the optimal
$t'_v$ are chosen to have the high density of states at Fermi surface. This
may be related to the observed extended region of flat band by
ARPES\cite{dagotto}.

\begin{figure}
\rotatebox{-90}{\includegraphics*[width=60mm]{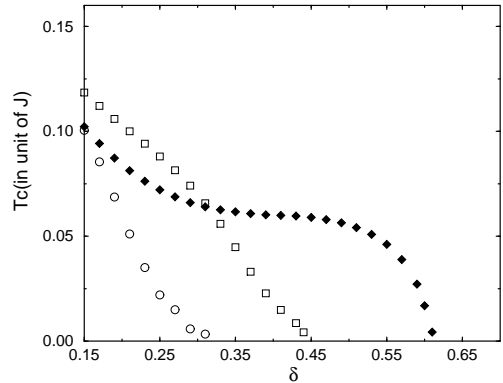}}
\caption{Mean-field results of $T_c$ for overdoped regime with
$(t',t'')=(0,0)$ (circles), $(-0.3,0)$ (squares) and $(-0.3,0.15)$
(diamonds).} \label{f:mft1}
\end{figure}

The results of  slave-boson MF calculation\cite{ubben}
 in Fig.\ref{f:mft1} show
similar behavior for the overdoped regime that indeed $t'$
enhances $T_c$. Including $t''$ pushes the superconducting regime
to even larger doping density $\delta$ by occupying the momenta
around $(\pi,0)$. A similar effect of $t''$ can also be seen in
Fig.\ref{f:phase_144v} by comparing the $(t',t'')=(-0.3,0)$ (full
squares) and $(-0.3,0.15)$ (open triangles) curves. Since the
slave-boson method is not quite reliable quantitatively in the
underdoped regime, we did not show the values of $T_c$ in
Fig.\ref{f:mft1}. However, if we do take the values literally, the
values achieved for $T_{c,max}$ are not as greatly enhanced by
$t'$ as for the VMC result shown in Fig.\ref{f:144t2}.
Similar results are reported by
the interlayer tunneling model\cite{hqlin2003}.

\begin{figure}[here]
\rotatebox{-90}{\includegraphics[width=2.5in]{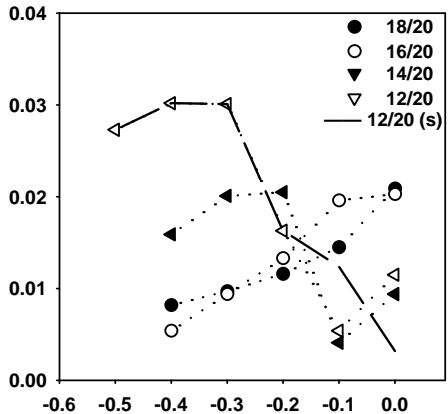}}
\caption{$P_d({\bf R}=(1,3))$ for versus $t'$ ($t''=-t'/2$) for
different 2 (open circle) , 4 (full circle), and 6 (open triangle)
holes in 20 sites. The dash line shows $P_d$ for the same symmetry
of the 8-hole case.} \label{f:ed20}
\end{figure}

Fig.\ref{f:ed20} shows the pair-pair correlation\cite{pair_ed} for
the longest distance ${\bf R}=(1,3)$ for 20-site lattices obtained
by the ED method. Pairing correlations for 2 and 4 holes are
suppressed by $t'$ and $t''$, but enhanced for the overdoped 6-
and 8-hole cases. The non-monotonic behavior of the overdoped
cases  is due to the level crossing of this system.
If we focus on the $s-$like symmetry states,
\pd will vary monotonically in the region $0\ge t'\ge -0.3$. The
result of ED method is quite consistent with the variational and
MF results that the enhancement of \pd by $t'$ occurs for larger
hole densities.

In summary, in the optimal and overdoped regions, SC is greatly
enhanced because of the deformation of the Fermi surface at these
doping densities. \npi is enhanced by including $t'$ and $t''$.
The occupation of $(\pi,0)$ by electrons is important for the
enhancement. The maximum enhancement of pairing correlation seems
to be reached for $-t'/t=0.3\sim 0.4$. On the other hand, \pd is
not enhanced for the underdoped regime as $n({\bf k}=(\pi,0))$ is
hardly affected by including $t'$.
It is well accepted that the physics on the overdoped side
is apparently much simpler in
that
experiments on the cuprates and theory for the t-J model indicate
that the overdoped materials are very close to ordinary Fermi liquids, whence
`Fermi-surface-based' arguments like ours are much more reliable than
in the underdoped region.
Our result shows that the
extended $t-J$ model naturally predicts the strong
correlation\cite{pavarini01} between $t'/t$ and $T_{c,max}$
observed in experiments for monolayer cuprates. In addition it
also indicates that further increase of $t'$ beyond what mercury
cuprates have most likely will not enhance $T_{c,max}$.

Although we have consistent results from VMC, ED methods and
slave-boson MF calculations, the optimal doping density,
$\delta_{opt}$, in Fig.\ref{f:phase_144v} is around 0.3 instead of
0.17 obtained in the experiments and in the VMC results without
including $t'$. But this is actually not a drawback. As argued in
References\cite{shih98, lee02}, VMC is expected to overestimate
the  values of the variational parameters $\Delta$ which is
related to the exhange energy $J$. Hence the pairing correlation
is definitely  much larger than that of a real ground state. From
our previous experiences the $\delta_{opt}$ seems to be always
shifted to a smaller value when we improve the variational wave
functions.  Thus in the future work going beyond VMC calculations,
we believe there is a better chance that we will have
$\delta_{opt}$  closer to the experimental value. We also expect
interlayer coupling will be important in getting the correct
$\delta_{opt}$. Now $t'$ is shown to be important in
enhancing pairing and it is also present in all high $T_c$
cuprates, the debate\cite{lee02,sorella02b} about pairing
robustness in the 2D $t-J$ model without $t'$ becomes somewhat irrelevant.
The agreement of $\delta_{opt}$ between $t-J$ VMC and experiments
looks fortuitous.

One of the consequences of our results is that the shape
of the Fermi surface plays an important role for high
temerature superconductors in the optimal and overdoped regions.
Fig.\ref{f:para144} (b) and (c) show
that the Fermi surface changes from hole-like to electron-like
once the maximum pairing is reached and the pairing is reduced as
doping increases. This is consistent with the ARPES results for
$La_{2-x}Sr_xCuO_4$\cite{ino02}. It may also be related to the
recent experiment\cite{balakirev03} which shows that the
low-temperature Hall coefficient  for the
$Bi_2Sr_{1.51}La_{0.49}CuO_{6+\delta}$ system exhibits a sharp
change at the optimal doping density. Clearly this issue deserves
more detailed study in the future.

The work is supported by the National Science Council in Taiwan
with Grant nos. NSC-91-2112-M-029-005, 006, and 068, and
NSC-92-2112-M-007-038. Part of the calculations are performed in
the IBM P690 in the National Center for High-performance Computing
in Taiwan, and the PC clusters of the Department of Physics and
Department of Computer Science and Engineering of Tunghai
University, Taiwan. We are grateful for their support.



\end{document}